\documentclass[aps,prl,twocolumn,superscriptaddress]{revtex4}
\usepackage{graphicx}

\begin{document}


\title{First results on angular distributions of thermal dileptons in nuclear collisions}




\author{R.~Arnaldi}
\affiliation{Universit\`a di Torino and INFN,~Italy}
\author{K.~Banicz}
\affiliation{CERN, 1211 Geneva 23, Switzerland}
\affiliation{Physikalisches~Institut~der~Universit\"{a}t Heidelberg,~Germany}
\author{J.~Castor}
\affiliation{LPC, Universit\'e Blaise Pascal and CNRS-IN2P3, Clermont-Ferrand, France}
\author{B.~Chaurand}
\affiliation{LLR, Ecole Polytechnique and CNRS-IN2P3, Palaiseau, France}
\author{C.~Cical\`o}
\affiliation{Universit\`a di Cagliari and INFN, Cagliari, Italy}
\author{A.~Colla}
\affiliation{Universit\`a di Torino and INFN,~Italy}
\author{P.~Cortese}
\affiliation{Universit\`a di Torino and INFN,~Italy}
\author{S.~Damjanovic}
\affiliation{CERN, 1211 Geneva 23, Switzerland}
\affiliation{Physikalisches~Institut~der~Universit\"{a}t Heidelberg,~Germany}
\author{A.~David}
\affiliation{CERN, 1211 Geneva 23, Switzerland}
\affiliation{Instituto Superior T\'ecnico, Lisbon, Portugal}
\author{A.~de~Falco}
\affiliation{Universit\`a di Cagliari and INFN, Cagliari, Italy}
\author{A.~Devaux}
\affiliation{LPC, Universit\'e Blaise Pascal and CNRS-IN2P3, Clermont-Ferrand, France}
\author{L.~Ducroux}
\affiliation{IPN-Lyon, Universit\'e Claude Bernard Lyon-I and CNRS-IN2P3, Lyon, France}
\author{H.~En'yo}
\affiliation{RIKEN, Wako, Saitama, Japan}
\author{J.~Fargeix}
\affiliation{LPC, Universit\'e Blaise Pascal and CNRS-IN2P3, Clermont-Ferrand, France}
\author{A.~Ferretti}
\affiliation{Universit\`a di Torino and INFN,~Italy}
\author{M.~Floris}
\affiliation{Universit\`a di Cagliari and INFN, Cagliari, Italy}
\author{A.~F\"orster}
\affiliation{CERN, 1211 Geneva 23, Switzerland}
\author{P.~Force}
\affiliation{LPC, Universit\'e Blaise Pascal and CNRS-IN2P3, Clermont-Ferrand, France}
\author{N.~Guettet}
\affiliation{CERN, 1211 Geneva 23, Switzerland}
\affiliation{LPC, Universit\'e Blaise Pascal and CNRS-IN2P3, Clermont-Ferrand, France}
\author{A.~Guichard}
\affiliation{IPN-Lyon, Universit\'e Claude Bernard Lyon-I and CNRS-IN2P3, Lyon, France}
\author{H.~Gulkanian}
\affiliation{YerPhI, Yerevan Physics Institute, Yerevan, Armenia}
\author{J.~M.~Heuser}
\affiliation{RIKEN, Wako, Saitama, Japan}
\author{M.~Keil}
\affiliation{CERN, 1211 Geneva 23, Switzerland}
\affiliation{Instituto Superior T\'ecnico, Lisbon, Portugal}
\author{L.~Kluberg}
\affiliation{CERN, 1211 Geneva 23, Switzerland}
\affiliation{LLR, Ecole Polytechnique and CNRS-IN2P3, Palaiseau, France}
\author{C.~Louren\c{c}o}
\affiliation{CERN, 1211 Geneva 23, Switzerland}
\author{J.~Lozano}
\affiliation{Instituto Superior T\'ecnico, Lisbon, Portugal}
\author{F.~Manso}
\affiliation{LPC, Universit\'e Blaise Pascal and CNRS-IN2P3, Clermont-Ferrand, France}
\author{P.~Martins}
\affiliation{CERN, 1211 Geneva 23, Switzerland}
\affiliation{Instituto Superior T\'ecnico, Lisbon, Portugal}
\author{A.~Masoni}
\affiliation{Universit\`a di Cagliari and INFN, Cagliari, Italy}
\author{A.~Neves}
\affiliation{Instituto Superior T\'ecnico, Lisbon, Portugal}
\author{H.~Ohnishi}
\affiliation{RIKEN, Wako, Saitama, Japan}
\author{C.~Oppedisano}
\affiliation{Universit\`a di Torino and INFN,~Italy}
\author{P.~Parracho}
\affiliation{CERN, 1211 Geneva 23, Switzerland}
\author{P.~Pillot}
\affiliation{IPN-Lyon, Universit\'e Claude Bernard Lyon-I and CNRS-IN2P3, Lyon, France}
\author{T.~Poghosyan}
\affiliation{YerPhI, Yerevan Physics Institute, Yerevan, Armenia}
\author{G.~Puddu}
\affiliation{Universit\`a di Cagliari and INFN, Cagliari, Italy}
\author{E.~Radermacher}
\affiliation{CERN, 1211 Geneva 23, Switzerland}
\author{P.~Ramalhete}
\affiliation{CERN, 1211 Geneva 23, Switzerland}
\author{P.~Rosinsky}
\affiliation{CERN, 1211 Geneva 23, Switzerland}
\author{E.~Scomparin}
\affiliation{Universit\`a di Torino and INFN,~Italy}
\author{J.~Seixas}
\affiliation{Instituto Superior T\'ecnico, Lisbon, Portugal}
\author{S.~Serci}
\affiliation{Universit\`a di Cagliari and INFN, Cagliari, Italy}
\author{R.~Shahoyan}
\affiliation{CERN, 1211 Geneva 23, Switzerland}
\affiliation{Instituto Superior T\'ecnico, Lisbon, Portugal}
\author{P.~Sonderegger}
\affiliation{Instituto Superior T\'ecnico, Lisbon, Portugal}
\author{H.~J.~Specht}
\affiliation{Physikalisches~Institut~der~Universit\"{a}t Heidelberg,~Germany}
\author{R.~Tieulent}
\affiliation{IPN-Lyon, Universit\'e Claude Bernard Lyon-I and CNRS-IN2P3, Lyon, France}
\author{G.~Usai}
\affiliation{Universit\`a di Cagliari and INFN, Cagliari, Italy}
\author{R.~Veenhof}
\affiliation{CERN, 1211 Geneva 23, Switzerland}
\author{H.~K.~W\"ohri}
\affiliation{Universit\`a di Cagliari and INFN, Cagliari, Italy}
\affiliation{Instituto Superior T\'ecnico, Lisbon, Portugal}

\collaboration{NA60 Collaboration}\noaffiliation



\date{\today}

\begin{abstract}

The NA60 experiment at the CERN SPS has studied dimuon production in
158A GeV In-In collisions. The strong excess of pairs above the known
sources found in the complete mass region 0.2$<$M$<$2.6 GeV has
previously been interpreted as thermal radiation. We now present first
results on the associated angular distributions. Using the
Collins-Soper reference frame, the structure function parameters
$\lambda$, $\mu$ and $\nu$ are measured to be zero, and the projected
distributions in polar and azimuth angles are found to be uniform. The
absence of any polarization is consistent with the interpretation of
the excess dimuons as thermal radiation from a randomized system.

\end{abstract}

\maketitle


Lepton pairs are a particularly attractive observable to study the hot
and dense matter created in high-energy nuclear collisions. Their
continuous emission, undisturbed by final-state interactions, probes
the entire space-time evolution of the fireball, including the early
phases with the conjectured QCD phase transitions of {\it chiral
symmetry restoration} and {\it parton deconfinement}. To the extent
that the bulk constituents of the expanding matter (hadrons and
partons) equilibrate, the direct lepton pairs generated by them are
commonly referred to as 'thermal radiation'. Our previous work has
indicated 'thermal' dilepton production to be largely mediated for
M$<$1 GeV by $\pi^{+}\pi^{-}$ annihilation via the strongly broadened
vector meson $\rho$~\cite{Arnaldi:2006jq}, and for M$>$1 GeV by
partonic processes like $q\bar{q}$
annihilation~\cite{Arnaldi:2007ru,Shahoyan:2008ejc}. The two dilepton
variables basically explored in this work were mass M and transverse
momentum p$_{T}$, where the correlations between the two were decisive
in bearing out the nature of the emission sources in the two mass
regions~~\cite{Arnaldi:2007ru,Shahoyan:2008ejc}.

Further information on the production mechanism and the distribution
of the annihilating particles, complementary to that from M and
p$_{T}$, can be obtained from the study of dilepton angular
distributions. This Letter presents the first measurement of full
dilepton angular distributions in the field of high-energy nuclear
collisions. It is restricted to the mass region $M$$<$1 GeV, due to
the lack of sufficient statistics for M$>$1 GeV. The question asked is
simple: can one get direct experimental insight into whether the
radiating matter is {\it thermalized}?

Historically, the interest in angular distributions of continuum
lepton pairs was mostly motivated by the study of the Drell-Yan
mechanism, following in particular the insight that the 'na\"{i}ve'
QED interpretation~\cite{Drell-Yan:1970} had to be significantly
modified due to QCD
effects~\cite{CollinsSoper:1977,LamTung:1978,Badier:1981ti,Falciano:1986wk}. The
differential decay angular distribution in the rest frame of the
virtual photon with respect to a suitably chosen set of axes, ignoring
the rest mass of the leptons, can quite generally be written as
\begin{equation}
\label{eq1}
\frac{1}{\sigma}\frac{d\sigma}{d\Omega} \propto (1+\lambda
cos^{2}\theta + \mu sin2\theta cos\phi + \frac{\nu}{2}sin^{2}\theta
cos2\phi)
\end{equation}

The angular dependence on polar angle $\theta$ and azimuth angle
$\phi$ dates back to~\cite{Gottfried:1964nx}, but the specific
coefficients $\lambda$, $\mu$ and $\nu$, the 'structure function'
parameters, follow the nomenclature in
e.g.~\cite{Falciano:1986wk,Brandenburg:1993cj}. They are directly
related to the helicity structure functions $W_{i}$ defined
in~\cite{LamTung:1978}, and in particular to the spin density matrix
$R_{ij}^{\gamma^{*}}$ of the virtual photon $\gamma^{*}$, the main
object of the spin
analysis~\cite{Badier:1981ti,Brandenburg:1993cj,Nachtmann}. We have
chosen here the Collins-Soper (CS) reference
frame~\cite{CollinsSoper:1977}, where the quantization axis $\vec{z}$
is defined as the bisector between the beam and negative target
momenta $\vec{p}_{beam}$ and $-\vec{p}_{target}$, which define the
reaction plane. The polar angle $\theta$ is then the angle between the
momentum of the positive muon $\vec{p}_{\mu^{+}}$ and the $\vec{z}$
axis, which define the decay plane, while the azimuth angle $\phi$ is
the angle between the reaction and the decay planes. However, the
particular choice of the frame is not really relevant here, since the
determination of the full set of coefficients $\lambda$, $\mu$ and
$\nu$ allows to compute them in any other frame by a simple
transformation~\cite{Falciano:1986wk}. This would not apply if only
the $cos\theta$ distribution would be measured.

\begin{figure}[t!]
\begin{center}
\includegraphics*[width=0.38\textwidth]{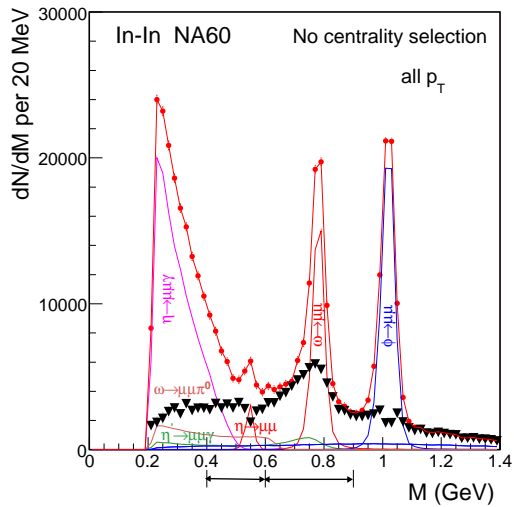}
\caption{Isolation of an excess above the electromagnetic decays of
neutral mesons (see text). Total data (closed circles), individual
cocktail sources (solid lines), difference data (thick triangles), sum
of cocktail sources and difference data (line through the closed
circles). Open charm still contained.}
   \label{fig1}
\end{center}
\vspace*{-0.8cm}
\end{figure}

In principle, the dilepton angular distributions can be anisotropic,
with all structure function parameters $\lambda, \mu, \nu \neq$
0~\cite{CollinsSoper:1977,LamTung:1978,Badier:1981ti,Falciano:1986wk,Brandenburg:1993cj}.
Even for spinless particles in the initial state like in
$\pi^{+}\pi^{-}$ annihilation, the parameters can still have any value
$\neq$ 0, since the spin density matrix of the virtual photon also
receives contributions from orbital angular
momentum~\cite{BratkovsToneev}. Very elementary examples are
$q\bar{q}$ and $\pi^{+}\pi^{-}$ annihilation along the beam direction
for p$_{T}$=0. Here $\mu$=$\nu$=0 and $\lambda$=+1 for $q\bar{q}$
(like lowest order DY~\cite{Drell-Yan:1970}) and $\lambda$=$-$1 for
$\pi^{+}\pi^{-}$~\cite{BratkovsToneev}, corresponding to transverse
and longitudinal polarization of the virtual photon, respectively.
However, a completely random orientation of annihilating partons or
pions in 3 dimensions (but not in 2~\cite{Nachtmann}) would lead to
$\lambda, \mu, \nu$ = 0~\cite{PHoyer,BratkovsToneev,Nachtmann}, and
that is the case of prime interest here.

Details of the NA60 apparatus are contained
in~\cite{Shahoyan:2008ejc,Banicz:2005nz}. The data sample for 158A GeV
In-In collisions is the same as used in~\cite{Arnaldi:2006jq,
Arnaldi:2007ru}, and the different analysis steps follow the same
sequence: assessment of the combinatorial background from $\pi$ and
$K$ decays by a mixed-event technique, assessment of the fake-matches
(associations of muons to non-muon vertex tracks in the Si pixel
telescope), isolation of the dimuon excess by subtraction of the known
meson decay sources and charm from the net opposite-sign sample, and
finally correction for acceptance and pair efficiency. All steps are
now done independently in each $[\frac{dN}{dcos\theta d\phi}]_{ij}$
bin. The binning is varied depending on the goal, thereby assuring
that the results are stable with respect to the bin widths chosen.

The assessment of the two background sources and open charm is
extensively discussed in the ref.~\cite{Shahoyan:2008ejc}. The
centrality-integrated net mass spectrum after background subtraction
is shown in Fig.~\ref{fig1} together with the contributions from
neutral meson decays: the 2-body decays of the $\eta$, $\omega$ and
$\phi$ resonances, and the Dalitz decays of the $\eta$, $\eta^{'}$ and
$\omega$. The data clearly exceed the sum of the decay sources. The
excess dimuons are isolated by subtracting them from the total (except
for the $\rho$), based solely on {\it local}
criteria~\cite{Arnaldi:2006jq, Arnaldi:2007ru}. The excess for M$<$1
GeV is interpreted as the strongly broadened $\rho$ which is
continuously regenerated by $\pi^{+}\pi^{-}$
annihilation~\cite{Arnaldi:2006jq, Arnaldi:2007ru}. Two adjacent mass
windows indicated in Fig.~\ref{fig1} are used for the subsequent
angular distribution analysis: the $\rho$-like region 0.6$<$M$<$0.9,
and the low-mass tail 0.4$<$M$<$0.6 GeV. To exclude the region of the
low-m$_{T}$ rise seen for all masses~\cite{Arnaldi:2007ru}, a
transverse momentum cut of p$_{T}$$>$0.6 GeV is applied, leaving about
54\,000 excess pairs in the two mass windows. The subtracted data for
the $\omega$ and $\phi$, about 73\,000 pairs, are subject to the same
further analysis steps as the excess data and are used for comparison.
\begin{figure}[b!]
\begin{center}
\includegraphics*[width=6.0cm, height=6.0cm]{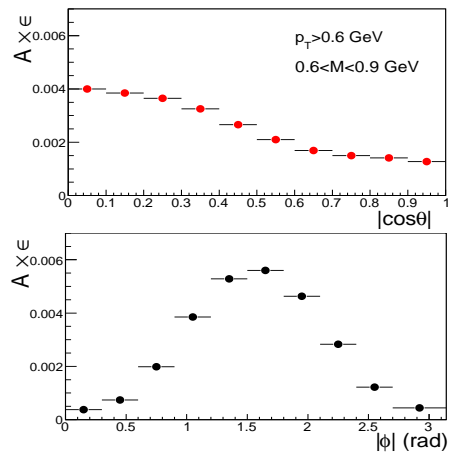}
\caption{Spectrometer acceptance as a function of the two angular
variables $|cos\theta|$ and $|\phi|$.}
   \label{fig2}
\end{center}
\end{figure}

\begin{figure*}[]
\begin{center}
\includegraphics*[width=0.33\textwidth]{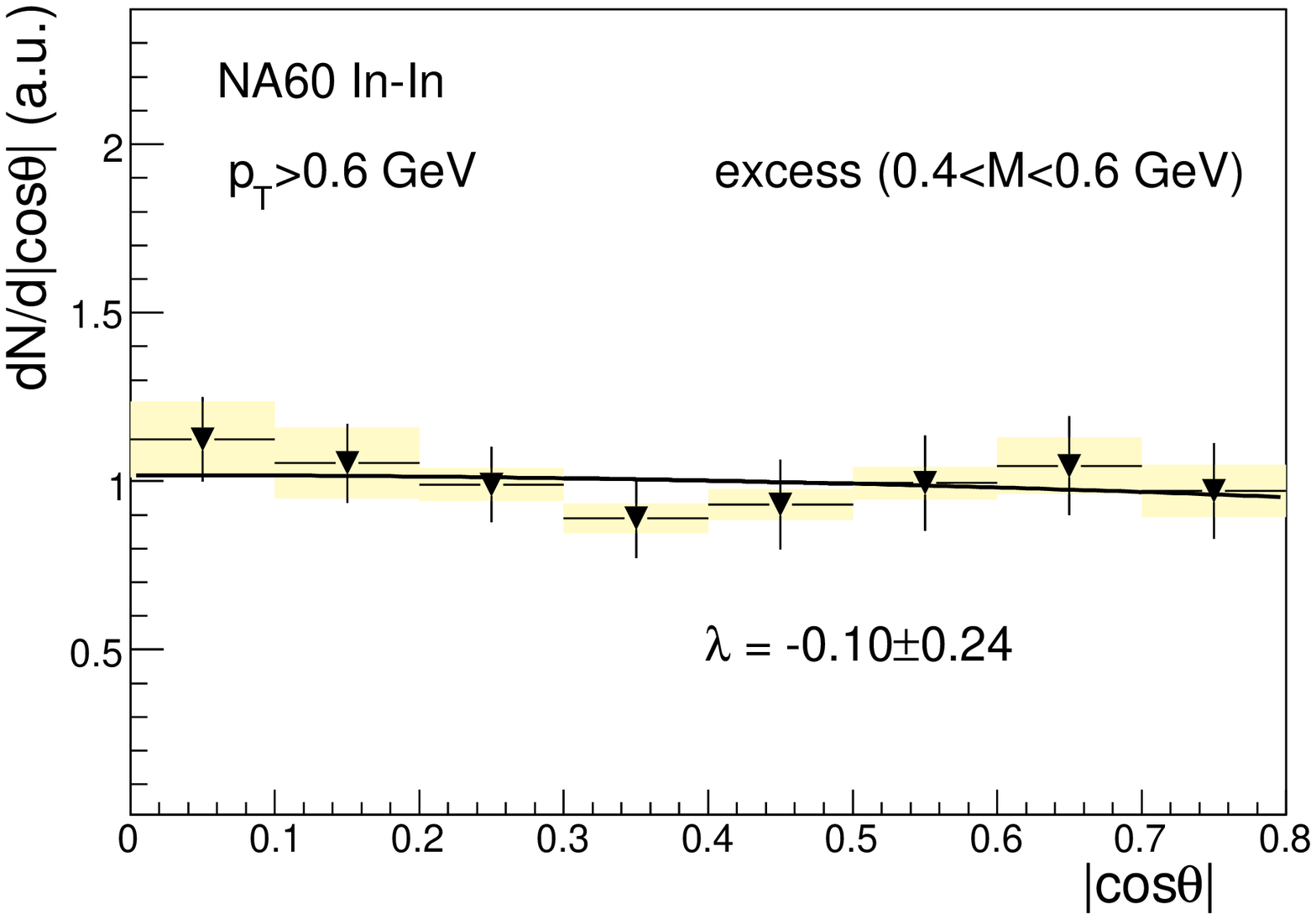}
\includegraphics*[width=0.33\textwidth]{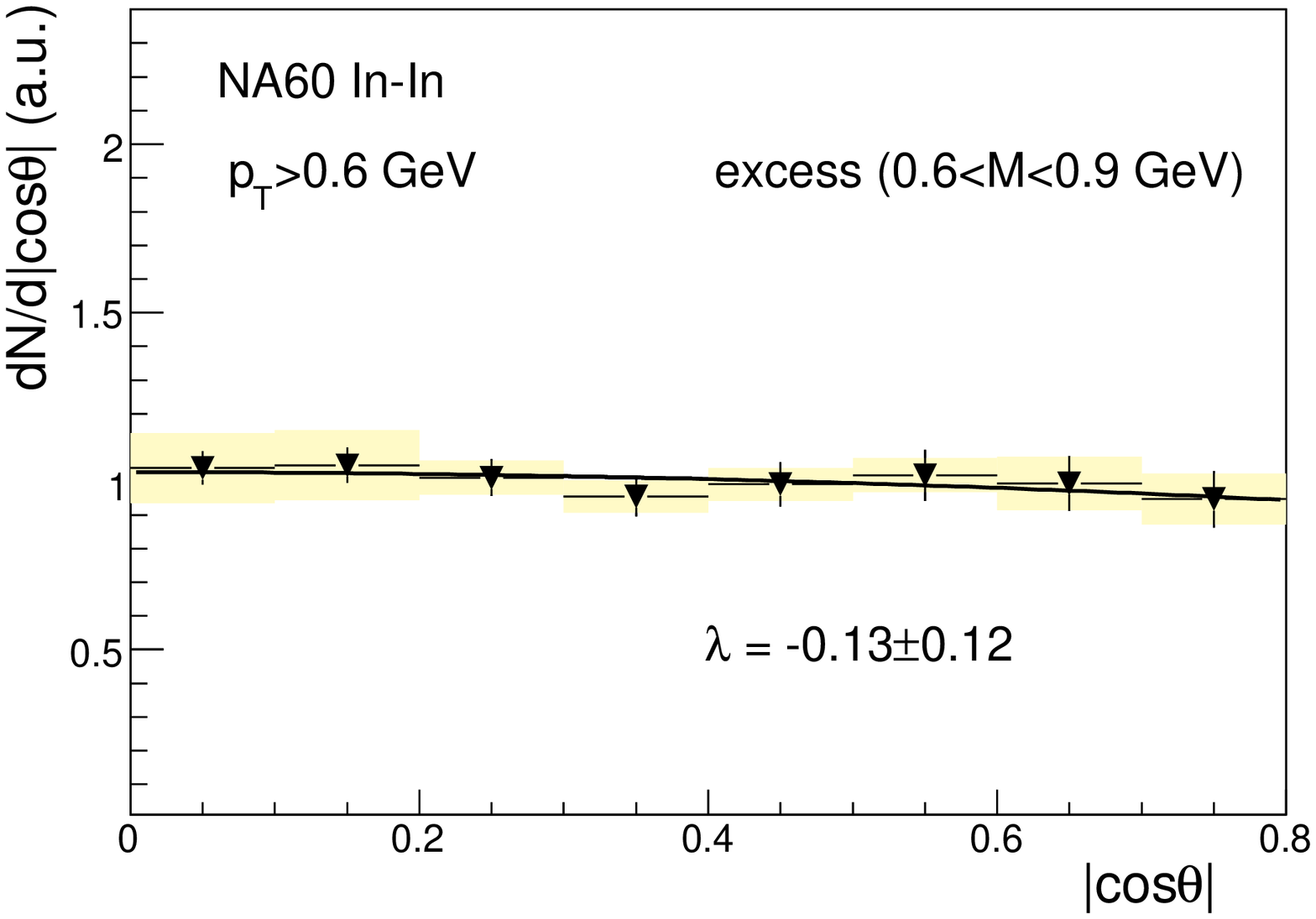}
\includegraphics*[width=0.33\textwidth]{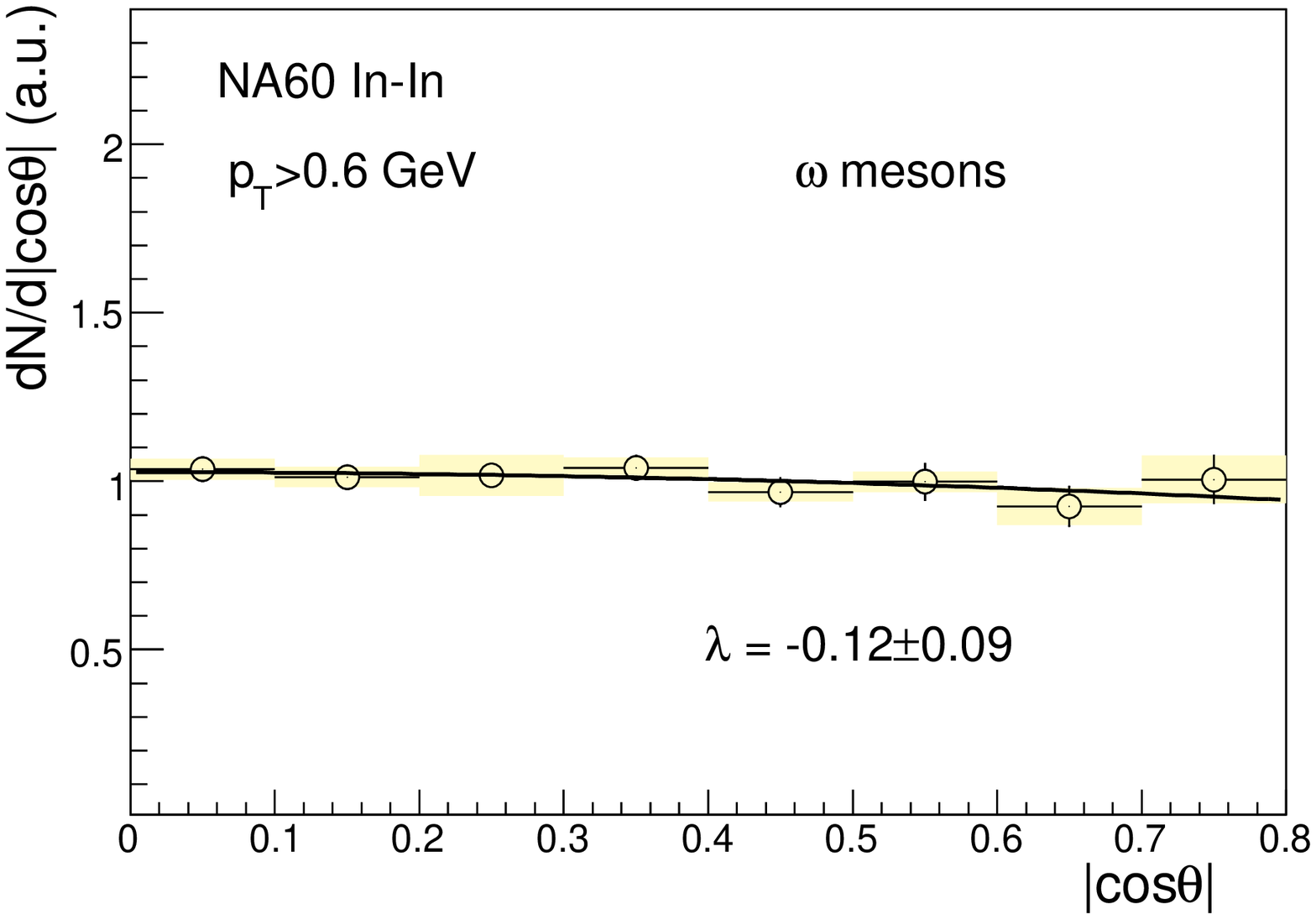}
\includegraphics*[width=0.33\textwidth]{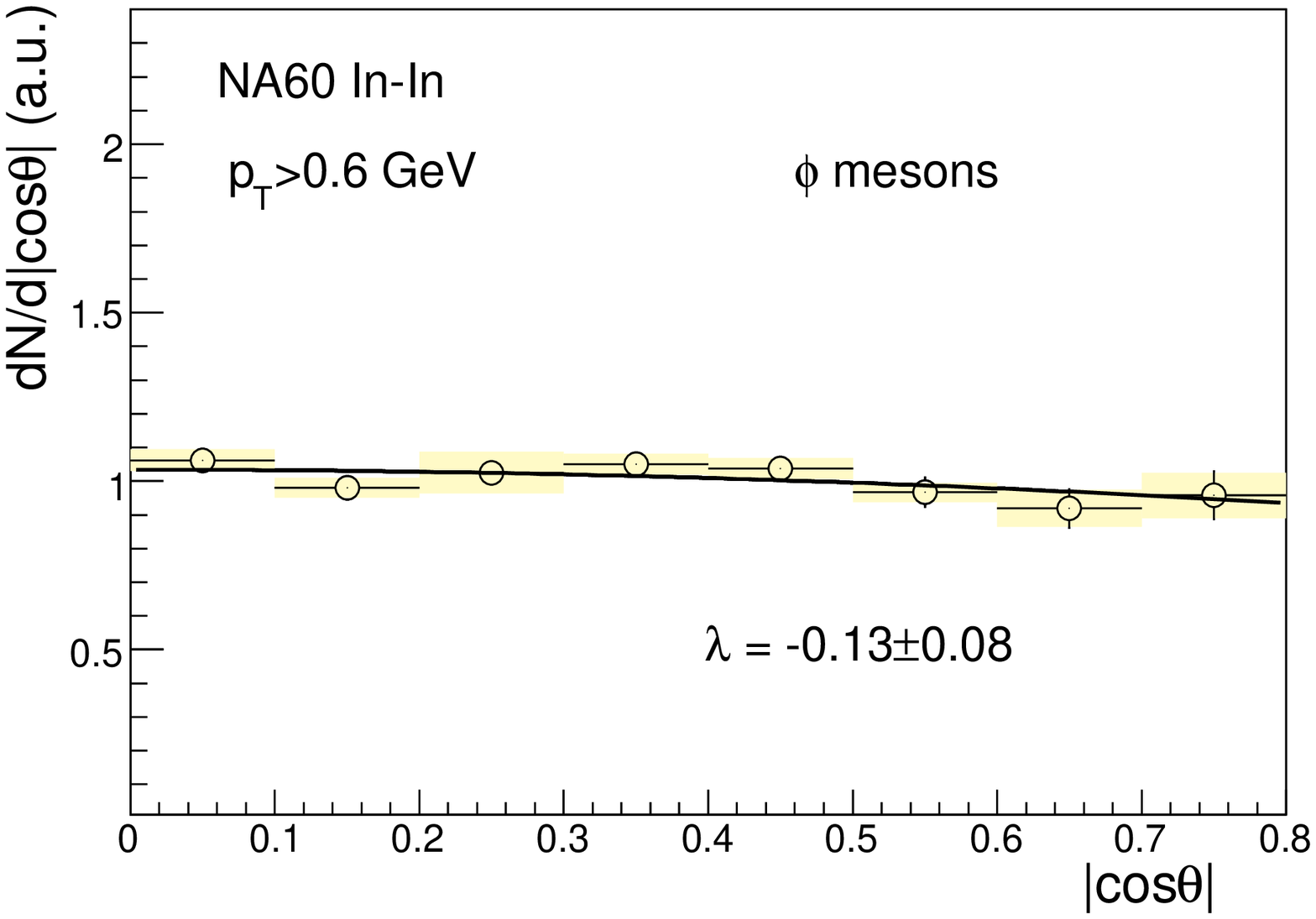}
\caption{Polar angle distributions of excess dileptons and of the
vector mesons $\omega$ and $\phi$.}
   \label{fig3}
\end{center}
\end{figure*}
The correction for the acceptance of the NA60 apparatus requires, in
principle, a 5-dimensional grid in M-p$_{T}$-$y$-$cos\theta$-$\phi$
space. To avoid large statistical errors in low-acceptance bins, the
correction is performed in 2-dim. $cos\theta$-$\phi$ space, using the
measured data for M, p$_{T}$~\cite{Arnaldi:2007ru} and
$y$~\cite{Damjanovic:2007qm} as an input to the Monte Carlo (MC)
simulation of the $cos\theta$-$\phi$ acceptance matrix. The
sensitivity of the final results to variations of this M-p$_{T}$-$y$
input has been checked, and the effects are found to be considerably
smaller than the statistical errors of the results. The MC simulations
were done in an overlay mode with real data to include the effects of
pair reconstruction efficiencies. The product
acceptance$\times$efficiency is illustrated in Fig.~\ref{fig2} for
0.6$<$M$<$0.9 GeV and p$_{T}$$>$0.6 GeV. The rapidity coverage is
3.2$<$y$<$4.2 (+0.3$<$y$_{cm}$$<$+1.3).

\begin{figure*}[]
\begin{center}
\includegraphics*[width=0.33\textwidth]{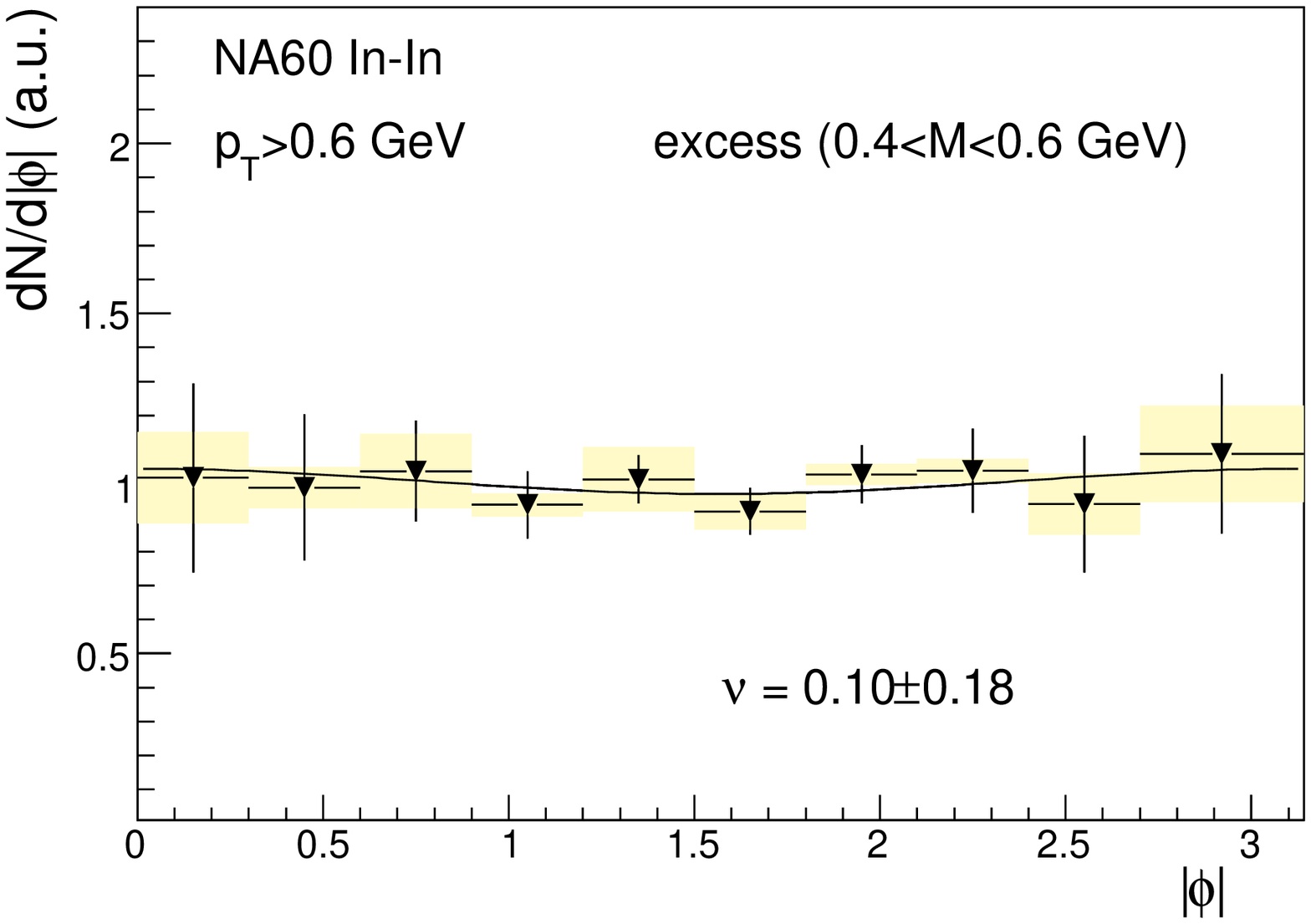}
\includegraphics*[width=0.33\textwidth]{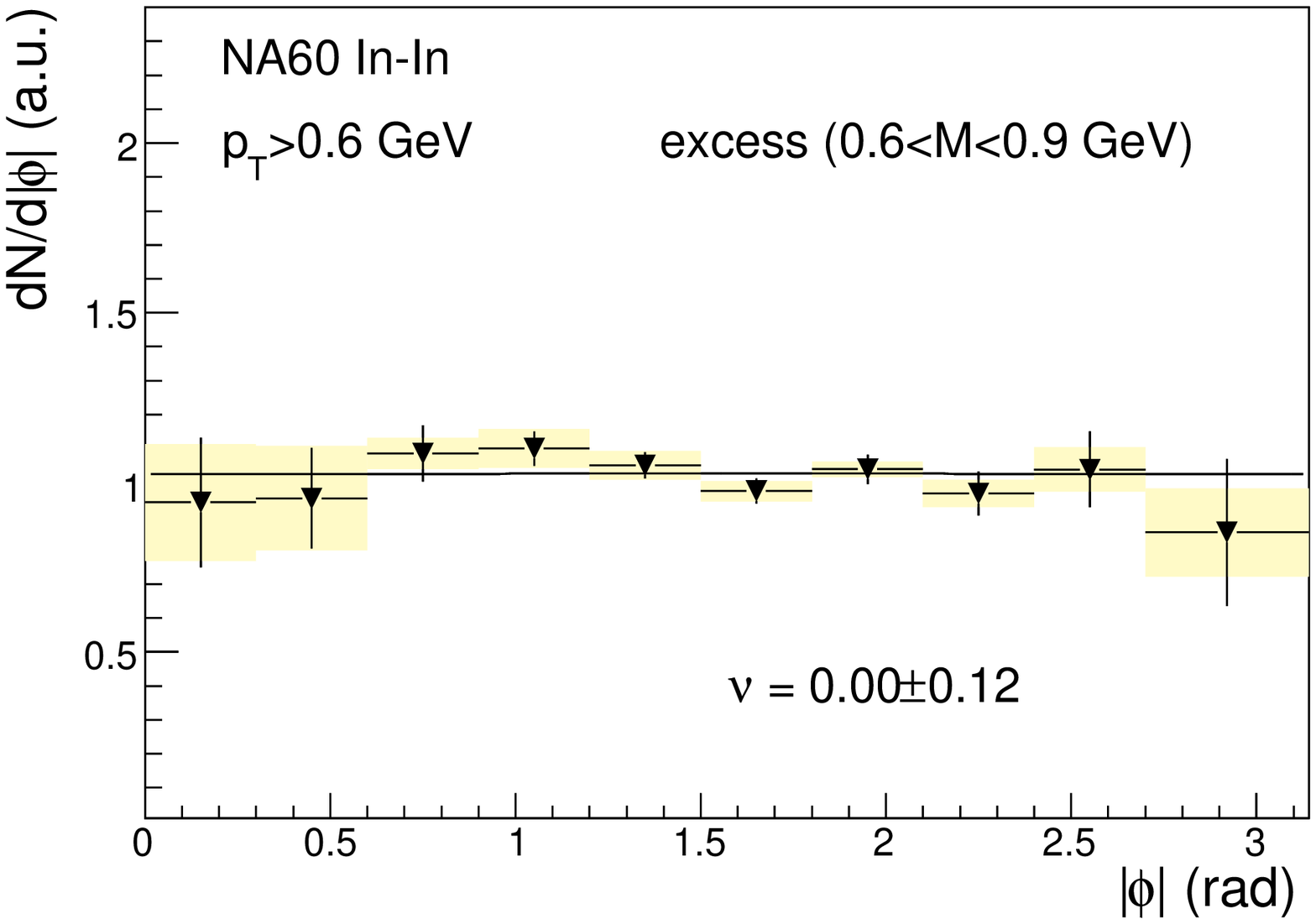}
\includegraphics*[width=0.33\textwidth]{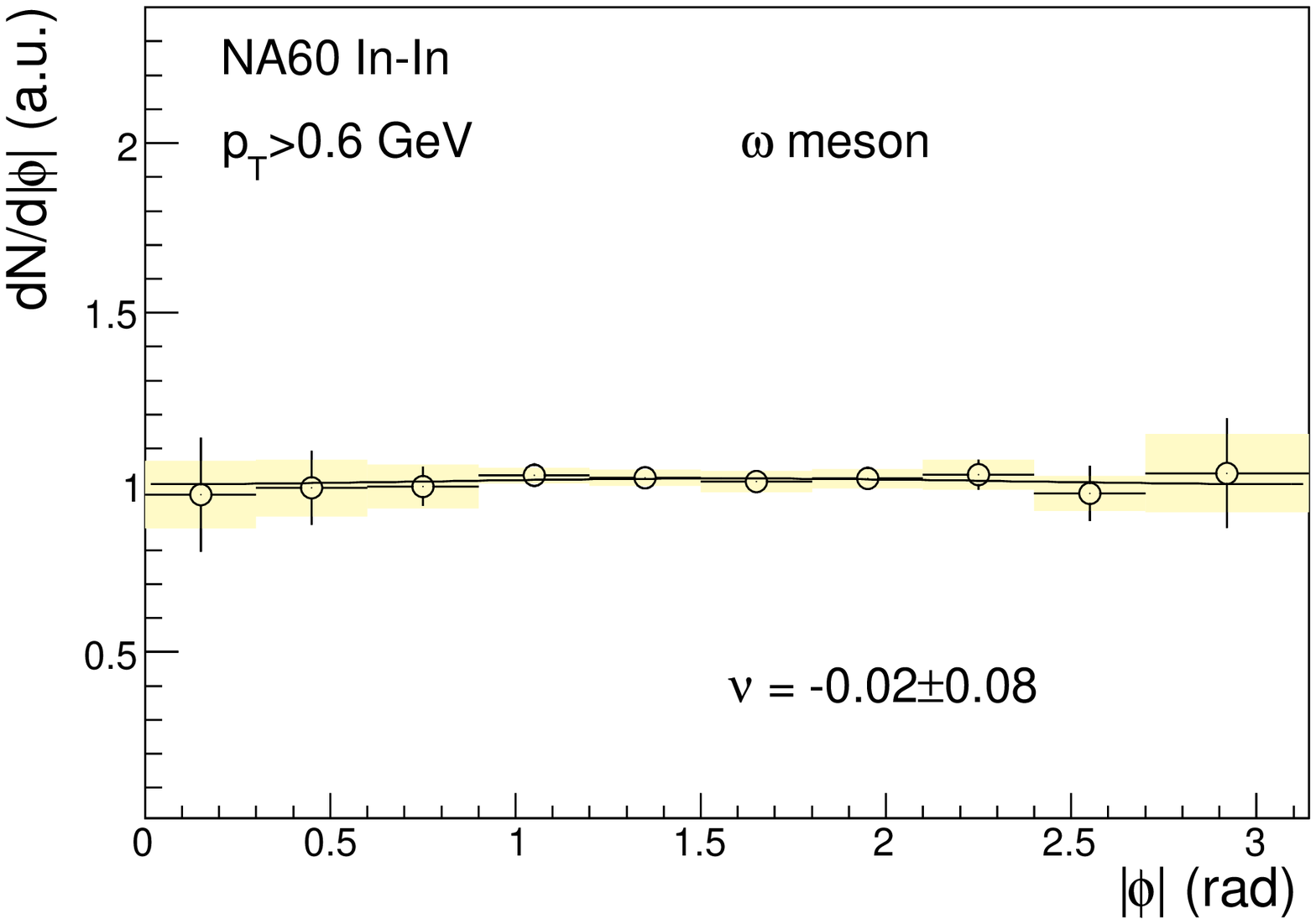}
\includegraphics*[width=0.33\textwidth]{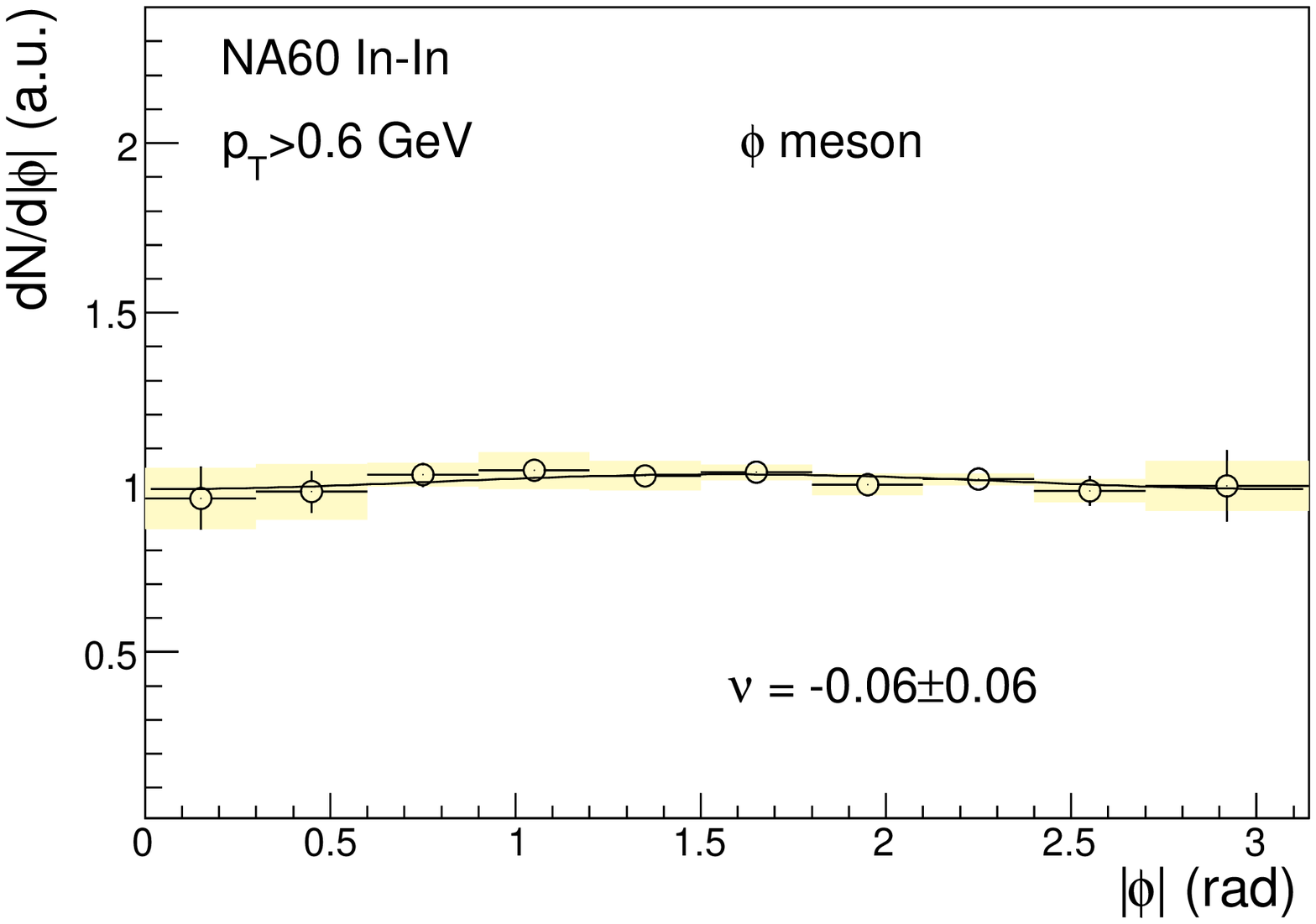}
\caption{Azimuth angle distributions of excess dileptons and of the
vector mesons $\omega$ and $\phi$.}
   \label{fig4}
\end{center}
\end{figure*}
The results on the angular distributions have been analysed in three
different ways, distinguished by the method and the associated
statistical/systematic errors. In the first and most rigorous method
(1), the 3 structure function parameters $\lambda$, $\mu$ and $\nu$
are extracted from a simultaneous fit of the 2-dimensional data on the
basis of~Eq.(\ref{eq1}), using a 6$\times$6 matrix in the range
$-$0.6$<$$cos\theta$$<$+0.6 (bin width 0.2) and
$-$0.75$<$$cos\phi$$<$0.75 (bin width 0.25). The restrictions in range
are enforced by regions of very low acceptance in the 2-dimensional
acceptance matrix, masked in the projections of Fig.~\ref{fig2}. The
fit values are summarized in Table~\ref{table1} for all 4 cases, the
two excess mass windows, the $\omega$ and the $\phi$. Within errors,
they are all compatible with zero. It is reassuring to see that this is also
true for $\mu$, as expected for a symmetric collision system at
midrapidity on the basis of symmetry considerations~\cite{Nachtmann}.

In the second method (2), setting now $\mu$=0, the 2-dimensional
acceptance-corrected data are projected onto either the $|cos\theta|$
or the $|\phi|$ axis, summing over the two signs. The polar angular
distribution is obtained by integrating~Eq.(\ref{eq1}) over the
azimuth angle ($\phi$)
\begin{equation}
\label{eq2}
\frac{dN}{dcos\theta} = A_{0}(1+\lambda cos^{2}\theta),
\end{equation}
while the azimuth angular distribution is obtained by integration over
the polar angle ($cos\theta$)
\begin{equation}
\label{eq3}
\frac{dN}{d\phi} = A_{1}(1+\frac{\lambda}{3}+\frac{\nu}{3}cos2\phi)
\end{equation}
The structure function parameters $\lambda$ and $\nu$ can then be
determined independently by 1-dimensional fits to the respective
projections. The data of the polar angular distributions together with
the fit lines according to~Eq.(\ref{eq2}) are shown in Fig.~\ref{fig3}
for all four cases, using now 8 bins in $|cos\theta|$ (bin width
0.1). The distributions are seen to be uniform, and the fit parameters
$\lambda$, included in Table~\ref{table1}, are again compatible with
zero, within errors. To determine the parameter $\nu$, $\lambda$
(contained in ~Eq.(\ref{eq3})) is set to the measured value of
$\lambda\pm\sigma_{\lambda}$. The fit results for $\nu$ on the basis
of~Eq.(\ref{eq3}), keeping the small number of bins used in method 1,
are again zero, within errors (see Table~\ref{table1}).

In the third method (3), the inclusive measured distributions in
$cos\theta$ and $\phi$ are analysed. A 1-dimensional acceptance
correction is applied in each case, determined by using (as now
measured) uniform distributions in $\phi$ (for $cos\theta$) and in
$cos\theta$ (for $\phi$) as an input to the MC simulations. The number
of bins in $|cos\theta|$ is kept, while that in $|\phi|$ is increased
to 10 (bin width 0.3). The data for the azimuth angular distributions
together with the fit lines according to~Eq.(\ref{eq3}) are shown in
Fig.~\ref{fig4}. The distributions are again uniform, as are those for
$cos\theta$ (not shown, since hardly distinguishable from
Fig.~\ref{fig3}). The resulting fit parameters for $\lambda$ and $\nu$
(taking account again of $\lambda\pm\sigma_{\lambda}$) are included in
Table~\ref{table1}. As expected, the errors are smaller than for the
other two methods, but the values of $\lambda$ and $\nu$ are still
close to zero, within errors.

\begin{table}[b!]
\caption{Summary of results for p$_{T}$$>$0.6 GeV on the structure
 function parameters $\lambda$, $\mu$ and $\nu$ in the CS frame,
 extracted from three different methods (see text). The $\chi^{2}/ndf$
 of the fits varies between 0.8 and 1.2. For a cut p$_{T}$$>$1.0 GeV,
 the results are the same, within errors.}
\label{table1}
\newcommand{\m}{\hphantom{$-$}}
\newcommand{\cc}[1]{\multicolumn{1}{c}{#1}}
\renewcommand{\tabcolsep}{0.1pc} 
\renewcommand{\arraystretch}{0.7} 
\begin{tabular}{@{}llll}
\hline
excess    & \cc{$\lambda$}    & \cc{$\nu$}        & \cc{$\mu$} \\
0.6$<$M$<$0.9 GeV &  & & \\
\hline
method1         &  \m-0.19$\pm$0.12 &  \m 0.03$\pm$0.15 &   \m 0.05$\pm$0.03 \\
method2         &  \m-0.13$\pm$0.12 &  \m 0.05$\pm$0.15 &   \m \\
method3         &  \m-0.15$\pm$0.09 &  \m 0.00$\pm$0.12 &   \m \\
\hline
\hline

\hline
excess    & \cc{$\lambda$}    & \cc{$\nu$} & \cc{$\mu$} \\
0.4$<$M$<$0.6 GeV &  &  &\\
\hline
method1        &  \m-0.13$\pm$0.27 &  \m 0.12$\pm$0.30 &   \m-0.04$\pm$0.10 \\
method2        &  \m-0.10$\pm$0.24 &  \m 0.11$\pm$0.30 &   \m \\
method3        &  \m-0.09$\pm$0.16 &  \m 0.10$\pm$0.18 &   \m \\
\hline
\hline

\hline
$\omega$ meson   & \cc{$\lambda$}    & \cc{$\nu$} & \cc{$\mu$} \\
\hline
method1        &  \m-0.10$\pm$0.10 &  \m-0.05$\pm$0.11 &   \m-0.05$\pm$0.02 \\
method2        &  \m-0.12$\pm$0.09 &  \m-0.06$\pm$0.10 &   \m \\
method3        &  \m-0.12$\pm$0.06 &  \m-0.02$\pm$0.08 &   \m \\
\hline
\hline

\hline
$\phi$  meson  & \cc{$\lambda$} & \cc{$\nu$} & \cc{$\mu$} \\
\hline
method1        &  \m-0.07$\pm$0.09 &  \m-0.10$\pm$0.08 &   \m 0.04$\pm$0.02 \\
method2        &  \m-0.13$\pm$0.08 &  \m-0.09$\pm$0.08 &   \m \\
method3        &  \m-0.05$\pm$0.06 &  \m-0.06$\pm$0.06 &   \m \\
\hline

\end{tabular}\\[2pt]
\end{table}

Figs.~\ref{fig3} and~\ref{fig4} also contain the systematic errors
attached to the individual data points. They mainly arise from two
sources. The subtraction of the combinatorial background, with
relative uncertainties of
1\%~\cite{Arnaldi:2006jq,Arnaldi:2007ru,Shahoyan:2008ejc}, leads to
errors of 2-3\% of the net data for the kinematic selection used
here. The subtraction of the meson decay sources causes (correlated)
errors for the excess and the vector mesons $\omega$ and $\phi$. With
respect to the excess, they range from 4-6\% up to 10-15\% in the
low-populated $cos\theta$-$\phi$ matrix bins. This variation is well
visible for the overall errors plotted in Figs.~\ref{fig3}
and~\ref{fig4}. Assuming, very conservatively, these errors to be
uncorrelated from point to point, the (statistical) fit errors for
$\lambda$ and $\nu$ quoted in Table~\ref{table1} would increase by
15-20\% if the systematic errors would be added in quadrature. Further
confidence into the stability of the results is obtained from their
independence of the methods and the bin widths used.

The global outcome from our analysis of dilepton angular distributions
is straightforward: the structure function parameters $\lambda$, $\mu$
and $\nu$ are all zero within the statistical and systematic errors,
and the projected distributions in polar and azimuth angle are all
uniform. This applies not only for the excess dileptons as anticipated
if of thermal origin, but also for the vector mesons $\omega$ and
$\phi$. While there may be a rather direct connection between the two
findings in nuclear collisions, it is of interest to note that the
result of $\lambda$=0 has been reported before for $\rho$ and $\omega$
production in $p$-$p$~\cite{Blobel} and $\pi^{-}$-$C$~\cite{Branson}.

We conclude, following the primary motivation of this study, that the
absence of any polarization is fully {\it consistent} with the
interpretation of the observed excess dimuons as {\it thermal
radiation} from a randomized system. While this is a necessary
condition, it is not sufficient. However, together with other features
like the Planck-like shape of the excess mass
spectra~\cite{Shahoyan:2008ejc,Damjanovic:2008ta}, the exponential
shape of the m$_{T}$ spectra~\cite{Arnaldi:2007ru,Damjanovic:2008ta}
and the global agreement with theoretical models both as to spectral
shapes and absolute yields~\cite{Shahoyan:2008ejc,Damjanovic:2008ta},
the thermal interpretation has become more plausible than ever before.

\begin{acknowledgments}
We are grateful to O.~Nachtmann for useful discussions. 
\end{acknowledgments}

\end{document}